**FINAL DRAFT**

Title: Observations of Solar Coronal Rain in Null-Point Topologies

Short Title: Observations of Coronal Rain in NPTs

E. I. Mason
Department of Physics, Catholic University of America, 620 Michigan Ave., N.E. Washington, DC 20064
04mason@cua.edu

Spiro K. Antiochos
Goddard Space Flight Center, 8800 Greenbelt Rd., Greenbelt, MD 20706
spiro.antiochos@nasa.gov

Nicholeen M. Viall
Goddard Space Flight Center, 8800 Greenbelt Rd., Greenbelt, MD 20706
nicholeen.m.viall@nasa.gov

ABSTRACT

Coronal rain is the well-known phenomenon in which hot plasma high in the Sun's corona undergoes rapid cooling (from ~ $10^6$ to < $10^4$ K), condenses, and falls to the surface. Coronal rain appears frequently in active region coronal loops and is very common in post-flare loops. This Letter presents discovery observations, which show that coronal rain is ubiquitous in the commonly occurring coronal magnetic topology of a large (~ 100 Mm scale) embedded bipole very near a coronal hole boundary. Our observed structures formed when the photospheric decay of active-region-leading-sunspots resulted in a large parasitic polarity embedded in a background unipolar region. We observe coronal rain to appear within the legs of closed loops well under the fan surface, as well as preferentially near separatrices of the resulting coronal topology: the spine lines, null point, and fan surface. We analyze 3 events using SDO Atmospheric Imaging Assembly (AIA) observations in the 304, 171, and 211 Å channels, as well as SDO Helioseismic and Magnetic Imager (HMI) magnetograms. The frequency of rain formation and the ease with which it is observed strongly suggests that this phenomenon is generally present in null-point topologies of this size scale. We argue that these rain events could be explained by the classic process of thermal nonequilibrium or via interchange reconnection at the null; it is also possible that both mechanisms are present. Further studies with higher spatial resolution data and MHD simulations will be required to determine the exact mechanism(s).

Subject keywords: null-point topologies, coronal rain, coronal holes, coronal heating

1. INTRODUCTION

One of the most intriguing features of the Sun's corona is that cold (T < $10^4$ K), dense (n > $10^{10}$ cm$^{-3}$) condensations frequently appear within the hot (T > $10^6$ K) corona. This phenomenon generally takes on two distinct forms: prominences/filaments and coronal rain. In prominences the cool mass can appear to sit stably in the corona for many hours, supported presumably by magnetic field lines with a dipped geometry (e.g., Antiochos et al. 1994). On the other hand, in coronal rain the condensations appear in the arched geometry of a coronal loop, in which case the condensations fall down the loop legs at speeds that are a significant fraction of the free fall velocity.

Coronal rain condensations are generally measured to have a density ~ $10^{10}$ - $10^{11}$ cm$^{-3}$ and a falling velocity ~ 50-100 km s$^{-1}$; however, there is compelling evidence that the condensations are highly structured in both temperature and density (Müller et al. 2005, Antolin et al. 2015). Observations have been made of the cooling material at many wavelengths since at least the 1970s, from extreme ultraviolet (EUV) to hydrogen alpha (H-alpha). For the purposes of this study, we adopt Antolin and Rouppe van der Voort's (2012) definition of coronal rain: "cool and dense blob-like material forming in the hot coronal environment in a timescale of minutes, which subsequently falls down to the surface along loop-like paths."

Coronal rain has long been observed in post-flare loops as in the classic "loop prominence systems" described by Bruzek (1964). For post-flare loops it seems clear that the condensations form as a consequence of the flare heating (e.g., Antiochos 1980). However, rain has also been widely reported in non-flare loops and around the borders of active regions (Schrijver 2001; O'Shea, Banerjee, & Doyle 2007; Ahn et al. 2014). In a post-flare loop, condensations appear and fall only once, but active region loops can undergo many cycles of coronal rain (Kawaguchi 1970, Antolin et al. 2015, Scullion et al. 2016, Auchère et al. 2018).

Two classes of models have been proposed for the origin of coronal rain, and of coronal condensations in general. The first, and most intuitive, is that the rain forms as a consequence of an abrupt cutoff in the heating, as expected for flares. The basic idea is that under certain widely expected conditions, flare plasma can cool down to H-alpha temperatures before it falls to the chromosphere (e.g., Antiochos 1980). Many observations appear to support this model (Scullion et al. 2014, Liu et al. 2015, Scullion et al. 2016). The second alternative model, referred to as "thermal non-equilibrium" (TNE), is far from intuitive, and was first proposed as the origin of the long-lived prominence condensations (Antiochos and Klimchuk 1991). The basic idea of TNE is that if the coronal heating is sufficiently localized near the chromospheric footpoints of a coronal loop (heating scale length ~10% of the loop length), then plasma at the loop midpoint cannot achieve a thermal equilibrium, even for steady heating. The loop plasma undergoes a quasi-cyclic evolution of heating and chromospheric evaporation followed by catastrophic radiative cooling resulting in condensation formation and fall (Antolin et al. 2015, Froment et al. 2015, Froment et al. 2017). Detailed numerical simulations have shown that as long as the heating is strongly localized near the chromosphere, TNE is a robust process for rain formation under very general loop geometries and heating properties (Mok et al 1990; Antiochos & Klimchuk 1991; Schrijver 2001; Müller et al. 2005; O'Shea, Banerjee, & Doyle 2007; Klimchuk, Karpen, & Antiochos 2010; Froment et al. 2015).

It is clear from the discussion above that observations of coronal rain provide unique constraints on the temporal and spatial properties of coronal heating. Since understanding coronal heating is, arguably, the most important unsolved problem in solar physics, detailed measurements of coronal rain are critically important. This letter reports the first observations of coronal rain in the well-known magnetic topology of an embedded bipole (Antiochos 1990; Lau and Finn 1990). Since these topologies are characterized by a 3D null point, we name these observed structures Raining Null Point Topologies (RNPT). Three example events are presented using SDO AIA and HMI data. These events exhibit rain for a range of time spans, some of which appears to be quasi-cyclic, and from several locations in the RNPT. Specifics of the observations will be introduced in the following section, followed by a discussion of the events and the implications of this discovery. One major new result from the observations is that so-called interchange reconnection (Crooker et al 2002) may be responsible for the coronal rain which falls along the spine and fan surface, as reported for some closed loops recently by Li et al. (2018). If so, it would be a new mechanism for rain formation. This is discussed in detail below.

## 2. OBSERVATIONS

### 2.1 Magnetic Field Properties

Embedded bipoles are a commonly observed magnetic structure for the solar magnetic field. They occur whenever an opposite-polarity region is embedded inside a larger background polarity at the photosphere. The resulting magnetic structure sketched in Figure 1a is defined by a single closed polarity inversion line (PIL) at the photosphere. In the corona, a dome-like surface (the separatrix) separates the background-polarity flux that connects across this PIL, from the background flux that is open or that connects to opposite polarities much further away. Somewhere on this dome, usually near the apex, a true 3D magnetic null point is present with two "spine" lines emanating from this null; one connects down inside the PIL and the other connects far away or is open (e.g., Antiochos 1998). Numerous observations starting with the "fountain regions" discovered by Skylab (Tousey et al. 1973; Sheeley et al. 1975) and continuing to the Yohkoh "anemones" (Shibata et al. 1994; Vourlidas & Bastian 1996) have verified that null-point topologies (NPT) are a generic feature of the solar atmosphere. Furthermore, a vast number of photospheric field extrapolations have demonstrated that NPTs are ubiquitous on a broad range of scales, from < 1 Mm to > $10^3$ Mm, and occur in active regions, quiet Sun, and coronal holes (e.g., Aulanier et al. 2000; Fletcher et al. 2001; Luhmann et al. 2003; Ugarte-Urra et al. 2007; Parnell & Edwards 2015). The SDO/AIA and SOHO/LASCO images of Figure 1b-d clearly show an example of this topology. For these RNPTs, all of the embedded bipoles have their origins as decaying leading sunspots of active regions.

NPTs are especially observable in coronal holes because the background flux that does not connect to the parasitic polarity region is open and dark in EUV/XUV, whereas the flux that connects into the parasitic polarity is closed and bright. Note that in this case the outer spine is open. A characteristic property of NPTs in coronal holes is that magnetic reconnection at the null and separatrix interchanges open and closed flux. This interchange reconnection occurs readily due to photospheric motions and has been postulated by many authors as the mechanism underlying both the plumes/rays and jets commonly observed in coronal holes (DeForest 1997, Raouafi et al. 2016, Shibata et al. 1992, Pariat et al. 2009, Wyper et al. 2017, etc.).

### Table 1: RNPT Characteristics

| RNPT | AR | Location | Null point height (Mm) | Average B (Gauss) | Continuous Rain (hours) |
|---|---|---|---|---|---|
| 2015/04/16 | SHARP 5437 | Bordering CH | 52 | 28 | 18 |
| 2015/05/04 | NOAA 12333 | Inside CH | 43 | 48 | 28 |
| 2016/02/27 | NOAA 12488 | Bordering CH | 112 | 21 | 59 |

The *raining null-point topologies* (RNPTs) that we describe in this Letter have at least two defining characteristics that we conjecture are sufficient for condensation formation to occur; these can be found summarized in Table 1. However, determining conclusively the necessary and sufficient conditions for RNPTs is beyond the scope of this paper due to the small number of observed events. We found the three events presented here during a manual search of SDO AIA 304 Å data from 2015 and 2016. Since finding these first RNPTs, we have extended the search back to 2011. The full time period has not yet been canvassed, but to date we have found 15 RNPTs.

The first condition is that they are of a size scale (~ 50-150 Mm) that is intermediate to the smaller, ~ 10 Mm scale commonly observed for jets and plumes (e.g., DeForest 1997; Raouafi et al. 2016) and the larger, ~ 0.5 Rs scale of true pseudostreamers (Wang 1994). Coronal rain has generally not been reported in the small closed structures of jets, and is not common in large pseudostreamers and streamers.

The second condition for RNPTs is that they occur near a coronal hole boundary, usually inside the hole. Figure 1e shows that the outer spine for this particular RNPT is open, hence, the RNPT must be in the coronal hole. This is a characteristic of jets, as well, since they form preferentially in coronal holes (Raouafi et al. 2008). Of course, pseudostreamers by definition are surrounded by open field. Every NPT of intermediate size that we have observed on an open/closed boundary does show rain. In contrast, there are many examples of small NPTs associated with coronal hole jets that do not show rain, and there are clear examples like the NPT of the Oct. 23, 2012 flare (Yang et al. 2015), which has the same size scale as our events but is far from a coronal hole, and never shows condensation formation. We conclude, therefore, that these two conditions are **sufficient** for observing coronal rain, at least, based on our limited number of events.

Related to this second condition is the formation process for RNPTs. The small parasitic polarities of jets are typically observed to form by photospheric flux emergence directly into a coronal hole; whereas, the large polarities of pseudostreamers typically are due to trailing polarity migrating into the vicinity of a coronal hole as a result of flux transport by surface motions and diffusion. Again, the RNPT formation process is somewhat intermediate to these two. The events we observed formed as a result of a bipolar active region emerging very near a coronal hole, and then the leading polarity diffusing outward as the active region decays. This formation process is somewhat unusual in that the leading sunspot had the opposite polarity of the main coronal hole in that hemisphere. The leading spot typically has the same polarity as the polar flux that produces the main coronal hole, but it is possible for the polarities to be opposite, especially late in the cycle when the emergence is near the equator. In fact, the three RNPTs that we describe below all emerged at latitudes equal to or less than 10˚. It is not clear whether this formation process is a necessary condition for condensation formation, or is simply one mechanism to produce NPTs of the correct size scale near coronal hole boundaries.

Another magnetic feature that RNPTs share with jets/pseudostreamers is the presence of a filament channel along their PIL, accompanied by observable cold prominence material in the longer-lived examples (i.e., Fig. 1c,d and Event 3, below). This is to be expected, given that RNPTs can persist for several Carrington rotations without major alterations to their structure (compare panels 4d-f, which show the same RNPT at three points across 25 days). It is well-known that filament channels will eventually form along any long-lived PIL (e.g., Mackay et al. 2018). Depending on the distribution of the photospheric flux, the PIL itself can be near-circular, as seen in Fig. 1c, but we also show other cases below where the PIL is simply a complex closed curve (see Event 2, below). In fact, for these cases the coronal field may well have a true pseudostreamer topology with multiple null points involving coronal hole corridors (Antiochos et al. 2011) or separator lines (Titov et al. 2011).

In summary, the magnetic properties of RNPTs are not distinctive. RNPTs appear to have the usual magnetic structure expected for a large jet or small pseudostreamer. However, their plasma properties, which we describe directly below, are clearly distinct.

*2.2 Plasma Dynamics*

*2.2.1 NOAA Active Region 12318*

We identified a small active region emerging from the eastern limb (Event 1) on 2015 April 5 near NOAA active region 12318. Labeled HMI SHARP 5437 (Space-Weather HMI Active Region Patch, see Bobra et al. 2014 and http://jsoc.stanford.edu/HMI/HARPS.html for the data product), it seems to have developed during the far-side portion of Carrington rotation 2162. STEREO-A data was not available for this event due to its superior conjunction period from March to July 2015, so the actual emergence date of AR 12318 is unknown. The small active region did not last very long, and had already decayed sufficiently to appear as an RNPT on the western limb of the Sun on April 16.

Figure 2a shows a context image of the Sun on April 11, with a PFSS model overlaid to show the southward extension of the northern polar coronal hole, and the fields of view of the next two panels outlined in black. Figures 2b and 2c present magnetogram and coronal imagery (HMI and AIA 211Å, respectively) on April 11, when HMI SHARP 5437 was near disk center. In keeping with the observational definition of coronal holes (Cranmer 2009), the lack of EUV signature in the 211 Å image shows that the RNPT sits very close to the open/closed boundary of the coronal hole to its north. The RNPT's negative polarity core is completely surrounded by positive polarity from another active region on its northwestern border, a coronal hole on its northern border, the remains of its own following sunspot on its eastern border, and weak plage on its western and southern borders. The coronal hole can be seen as the dark purple area extending from the upper left corner of the image in Figure 2c, while closed loops are in light purple to white. Figure 2d shows a time step from the period that Event 1 appeared on the limb, raining, during April 16. For this and the other two events described below, HMI data are shown while the associated AR was near disk-center in order to acquire the most detailed and accurate view possible of the photospheric magnetic field. The rain is shown off-limb in all cases so that the rain signal is not swamped by chromospheric emission.

Figure 2d clearly shows streams of falling coronal rain blobs during a squall that lasted for approximately 18 hours (see Movie 1, online; for a similar "shower," see Antolin & van der Voort 2012). In this event, most of the rain appears to be forming along the bottom of the outer spine and the longest legs of the dome. Event 1 apparently decays during its transit of the far side, as it is no longer in evidence during the next Carrington rotation. A movie of all the events may be found online.

*2.2.2 NOAA Active Region 12333*

NOAA active region 12333 was identified first on 2015 April 23, when it emerged over the eastern limb of the Sun already well developed. Figure 3a shows the general magnetic configuration of the Sun's northern hemisphere on April 28[th]; the extension of the northern polar coronal hole can be seen extending down along the side and wrapping all the way around the RNPT in this image. This RNPT's (Event 2) magnetic field was more elliptical than the others, and so it appears to possess more than one external spine, much as a true pseudostreamer possesses a fan extending from its null points (Figure 3b,c). Event 2 is situated in the middle of the southern extremity of a northern polar coronal hole, as evidenced by the uniformly positive magnetic field surrounding the parasitic polarity in Figure 3b, a paucity of signal in the EUV in Figure 3c, and the upflows shown by the EIS rasters in Figure 3d.

AR 12333 was also observed several days before reaching the western limb by the EUV Imaging Spectrometer (EIS) instrument on board Hinode. A 60-step raster observation ran from 23:02 on April 29 to 00:04 UTC on April 30, using 25 EUV lines spanning the temperature range from 0.2-17.8 MK; the yellow box in Figure 3a contextualizes the observation field of view. A sampling of the clearest images from the series can be seen in Figure 3d, showing the velocity shifts of plasma between 0.5-2 MK in temperature. All of these images show the well-known active region pattern of blueshifts around the perimeter of the closed-field region and redshifts within it (Harra et al. 2008, Kayshap et al. 2015). This supports the argument that RNPTs are isolated structures in magnetic connectivity, forming on or constituting open/closed boundaries.

Event 2 can be seen raining as soon as the top of the dome appears on the limb (around 03:00 UTC on May 3), and rains in several squalls that last tens of hours until it disappears off the limb on May 6 (Figure 3e and Movie 1, online). It is unclear at this point whether there is any periodicity present, such as would be expected from TNE, and such a search is beyond the scope of the present work. However, it would be an important objective for future studies of these events.

*2.2.3 NOAA Active Region 12488*

A compact but strong sunspot grouping was designated NOAA 12488 on 2016 January 20 around 10° N of the solar equator, and proceeded to grow rapidly during its disk transit. STEREO-A was at an angle of 164° with respect to the Earth, so the active region could easily be tracked across the far side of the Sun, where it underwent several small eruptions and grew a well-developed and roughly circular prominence surrounding the leading sunspot. The context image showing the RNPT developing on the border of the northern polar coronal hole is shown in Figure 4a. Note that the leading sunspot structure is still fairly robust as seen in Figure 4b, but the RNPT has already formed (Figure 4c) and rained continuously for at least 59 hours by this point (see Figure 4d and Movie 1). Renamed NOAA active region 12501, and here Event 3, it lasted for another solar rotation in this geometry.

It should be noted that, while it is difficult to determine whether the rain continued after the RNPT was completely on-disk, it was still raining when it entered the off-limb region on February 27 (Figure 4e), albeit far fewer blobs fell at a much lower frequency. This trend towards more infrequent rain is continued in the imagery of Figure 4f, which occurs once the RNPT re-emerges on the east limb on March 11. Figure 5 directs attention to another important facet of this particular observation: coronal rain is also visible precipitating far down the loop legs as distinct intensifications in the 171 Å observations. While it is known that coronal rain is highly multithermal and there are often loop intensifications in 171 during coronal rain events (Antolin et al., 2015), observations of distinct blobs in this channel are not often reported. This indicates that at least some of the coronal rain in RNPTs forms under different conditions, perhaps condensing faster or at a higher temperature than the rain seen in quiescent or flare-driven events.

*2.3 Quantitative Properties*

These three RNPTs encompass areas on the photospheric surface of roughly 2000, 2500, and $10^4$ square megameters, respectively (determined by the approximate location of the PIL in HMI observations of the active region, taken near disk center). The strongest magnetic signatures average about 1 kG, for both polarities; for the three events covered by this paper, the average was 1100 G for the positive polarity and -1300 for the negative. However, the average magnetic field measurements in the core and surrounding opposite polarity are in the low tens of gauss (~30 G and -35 G, respectively).

The rain in all of the cases presented here show the same general pattern of blob appearance, precipitation, and elongation (see Movie 1, online), common to other reports of coronal rain. A blob is detected near the loop top(s), which becomes much more elongated once the cascade down the loop leg begins. During storms the entire RNPT often appears almost completely outlined in rain, making analysis of specific blobs difficult; what follows are some brief remarks on the characteristics of rain blobs that were isolable.

Loop top blobs, when they can be observed clearly, tend to be roughly spherical and on the order of 5,000 km in radius. An analysis of 12 distinguishably unique rain units that fell along the legs from all three events discussed here are better characterized as streams rather than blobs, with an average detectable width in 304 Å of 2,600 km and an average length of 13,000 km; examples of both these geometries can be seen highlighted by arrows in Figure 3e. Most of the rain precipitates along one or both loop legs for extended periods, and the observed "streams" are likely a collection of loops generating smaller blobs being viewed along the same line of sight contemporaneously. It is therefore impossible to determine what the geometry of truly average rain blobs are in this context, due to the torrential nature of the "storms" and the limited resolution of SDO.

The velocity of the falling material also varies widely. Without a stereoscopic view of the domes, it is difficult to tell how rounded and symmetric the loop legs are, which complicates the determination of the actual distance that the rain is traveling. Other authors have found that the average speed of coronal rain in post-flare arcades tends to fall in the range of 50-100 km s$^{-1}$; our findings average 49 km s$^{-1}$ (Antolin et al. 2015). Discussion of possible reasons this number falls near the low end of previous findings will be pursued in the next section.

## 3. DISCUSSION

The observations presented above lead to several important conclusions. Most important is the discovery that NPTs associated with coronal holes and of intermediate scale (~ 100 Mm) frequently – and perhaps always – exhibit coronal rain. As detailed below, this result has strong implications for coronal heating and/or dynamics. In the previous sections we presented observations of coronal rain in RNPTs on three dates between 2015 and 2016. The ease with which these structures were identified and the frequency of rain during all observations provides compelling support for the conclusion that this is a ubiquitous phenomenon. Note that we have identified more RNPTs than the three presented in this letter (~ 15), and all those sharing the key conditions of proximity to a coronal hole and intermediate size scale rain heavily. The condensations form preferentially at the null and outer spine, but some form on large closed loops clearly inside the separatrix dome and null. In all cases the rain continues for days on what appears to be the same magnetic loops, so it is clearly not a one-shot phenomenon like flare cooling.

Another characteristic common to all RNPTs appears to be their magnetic origins. Like the NPTs of coronal hole jets, they form by flux emergence, as a leading polarity sunspot evolves near an opposite polarity open-field region, as the active region associated with the sunspot decays. This results in a relatively concentrated and strong parasitic polarity (> 1 KG) inside or very near a coronal hole. During the search for events, it became apparent that their distribution was not uniform across the solar cycle, but clustered around the ascending and descending phases. This is likely a result of Hale's Law, which states that active region leading sunspots tend to have the same polarity as their hemispheric polar region, which usually defines the polarity of the main coronal hole in that hemisphere. Consequently, it is only at certain times during the cycle that leading sunspots will appear with a polarity opposite to that of the main coronal hole and, thereby, have the possibility of forming a RNPT. The coronal topology of our events is generally that due to a single null point, but we have found at least one event that appears to have the multi-null-point topology of true pseudostreamers (Event 2). Furthermore, our events cover a broad range of geometries for the flux distribution and PIL at the photosphere. It appears, therefore, that the exact topology of the embedded bipole is not essential, only the size scale of the closed flux region and the local presence of open flux.

An additional important conclusion from the observations is that the mechanism driving the condensations appears to evolve during the AR decay phase. Several RNPTs can be seen raining continuously for days early in their formation, but "dry up" in observations upon arrival on the opposite limb without undergoing any other appreciable changes to its morphology or temperature range (see Figure 4d-f). The disappearance of condensations is most pronounced for those that form at the null and spine. This time dependence of the rain places important constraints on the possible mechanisms for its formation.

As discussed in the Introduction, two models have long been proposed for the origin of coronal rain: impulsive heating and TNE. The first seems unlikely to be relevant to our observations, because the rain persists for several days in apparently the same magnetic structures; whereas, a particular flare loop shows condensation formation only once as a result of the abrupt cessation of flare heating (e.g., Bruzek 1964, Antiochos 1980). There are, at least, two features in our results that support TNE as the rain mechanism. First, TNE produces quasi-cyclic squalls of rain, in agreement with the observations. Second, a basic property of TNE is that it favors condensation formation on the longest closed loops in the system (Antiochos & Klimchuk 1991), which in our case are those at or near the separatrix and null, again in agreement with the data. A further property of these loops is that they have a large expansion factor localized near the null; in principle, the expansion becomes infinite exactly at the null. We conjecture that such locations of large expansion near the loop apex are especially favorable for condensation formation, because a relatively large amount of coronal plasma will collect there. If the heating is localized near the chromosphere, as required for TNE, it would be difficult to support via thermal conduction the large radiative losses of the hot plasma in the region of large expansion high up in the corona, which would account for why condensations form preferentially near the null. This conjecture seems promising, but it needs rigorous testing with detailed numerical simulations, which we plan for a subsequent study.

Although it is likely that TNE is operating in RNPTs, there are at least two features in our observations that are in conflict with this mechanism and, therefore, imply that an additional process must be present. First, the rain, especially that associated with the null, decreases substantially as the RNPT ages. This is difficult to explain within the TNE model, because the size of the system does not change significantly. It could be argued that as the RNPT ages, the length scale for coronal heating becomes larger, but this explanation seems completely *ad hoc*. The second and even stronger conflict between the data and the TNE model is that for some events, condensations clearly form on the outer spine, which appears to be open. On open field lines, however, radiative losses are negligible compared to the enthalpy flux outward; consequently, TNE is not expected to occur. For example, coronal rain is generally not observed in the open flux of coronal holes.

We propose that interchange reconnection is the process responsible for the condensations along the outer spine and perhaps much of the rain associated with the null and separatrix. Interchange reconnection is not a new mechanism; it has long been proposed as the basic process driving plumes (DeForest et al. 1997) and jets (Pariat et al. 2009). To our knowledge it has not been proposed as a mechanism for condensation formation in this context. We claim, however, that it should naturally lead to cool condensations along the outer spine. As a result of interchange, a closed loop filled with hot, dense plasma reconnects at the null with an open flux tube containing cool, tenuous plasma. This results in a new open flux tube with a hot, dense lower section that form an arc along the separatrix dome and a tenuous upper section that forms the outer spine. There would be too much plasma in the lower section to be ejected as the solar wind, so most of it must fall back down to the chromosphere. If the interchange is fast, then the hot plasma will expand upward rapidly into the spine and cool down to chromospheric temperatures before falling to the chromosphere along the separatrix dome. Such an evolution would match our observations very well. Note that it naturally leads to condensations only along the outer spine and not on open flux tubes, in general. Furthermore, interchange could easily account for the decrease in rain activity as the RNPT ages. We expect the parasitic polarity to diffuse and weaken in time, which is likely to decrease the rate of interchange reconnection. If our conjectures are correct, then interchange reconnection may well be the primary mechanism for condensation formation in RNPTs, but again this model needs rigorous testing with numerical simulations.

In addition to possibly revealing a new mechanism for condensation formation, RNPTs may also have important implications for the broader coronal and solar wind. The location of RNPTs on the open/closed boundary makes them a likely observational target for remote missions, and for correlated *in situ* measurements. The slow solar wind is hypothesized to originate in this region, where interchange reconnection releases ion populations and magnetic field structures from both open and closed field lines into the heliosphere (e.g., Higginson et al 2017). RNPTs form on or in this region, often at the low latitudes most often sampled by *in situ* instruments; consequently, they could be detected and provide pivotal constraints on slow solar wind origins and dynamics. For example, it would be highly informative to measure if there are any correlations between the condensation formation observed in the corona and the plasma/magnetic structure in the corresponding slow wind. This could provide definitive proof that interchange reconnection is the mechanism driving both the slow wind and RNPTs.

RNPTs may also provide new insights into the nature of heating and magnetic reconnection in the corona. The NPTs of jets generally do not exhibit condensation formation, but this is not surprising given their small size scale, ~ 10 Mm or less. Neither TNE nor interchange reconnection is likely to have significant observable effects on very small closed loops. On the other hand, the result that streamers and pseudostreamers generally do not exhibit condensations is unexpected and places constraints on the mechanisms. The lack of coronal rain in very large null point topologies implies that there is some upper limit on the length scale for TNE and interchange reconnection to be effective at producing observable condensations; however, the physical origin for such a limit is far from clear.

A challenge to resolving the issues discussed above is the nature of the observations themselves. One uncertainty is the measurement of blob size and fall speed. Aside from the limited spatial resolution of instruments like AIA, there are line-of-sight difficulties. Tracking condensations through a loop leg by intensity signatures implicitly relies on the leading edge of the condensation remaining the same unit of plasma, which is by no means certain. Often condensations emitting in cool lines near the top of a leg change size and shape, possibly by accumulating cooling plasma along the leg, rather than by falling down it (Antolin et al. 2015). Overlapping, under-resolved loops and projection biases also add to this problem. These effects may be responsible for the somewhat low value, discussed above, that we measure for the condensation speeds. Doppler shifts measured in narrow passbands, such as those available from IRIS observations, may help resolve many of these challenges, but less event coverage is provided due to the smaller field of view. It is clear that much more observational and theoretical work remains to be done on these fascinating structures of Raining Null Point Topologies.

## 4. ACKNOWLEDGMENTS

The authors would like to thank Adrian Daw and Aleida Higginson for many insightful conversations during the development of this project, and their thoughtful comments on the draft. This work was supported by Adrian Daw and Douglas Rabin. NMV is supported by the Heliophysics Internal Scientist Funding Model. We would also like to acknowledge support from the NASA HSR and GI programs.

## 5. REFERENCES

Ahn, K., Chae, J., Cho, K., et al. 2014, SoPh, 289, 11

Antiochos, S. K. 1980, ApJ, 2, 4

Antiochos, S. K. 1990, MmSAI, 61

Antiochos, S. K., & Klimchuk, J. A. 1991, ApJ, 378

Antiochos, S. K. 1998, ApJL, 502

Antiochos, S. K., Mikić, Z., Titov, V. S., et al. 2011, ApJ, 731, 1

Antolin, P., & Rouppe Van Der Voort, L. 2012, ApJ, 745, 2

Antolin, P. et al. 2015, ApJ, 806, 1

Auchère, F., Froment, C., Soubrié, E., Antolin, P., Oliver, R., & Pelouze, G. 2018, ApJ 853, 2

Aulanier, G. et al. 2000, ApJ, 540

Bobra, M. G. et al. 2014, SoPh 289, 9

Bruzek, A. 1964, ApJ, 140, 2

Cranmer, S. R. 2009, LRSP, 6

Crooker, N. U., Gosling, J. T., & Kahler, S. W. 2002, JGR, 107, 1

DeForest, C. E., Hoeksema, J. T., Gurman, J. B., et al. 1997, SoPh, 175

Fletcher, L., Metcalf, T. R., Alexander, D., Brown, D. S., & Ryder, L. A. 2001, ApJ, 554, 451

Froment, C., Auchère, F., Bocchialini, K., et al. 2015, ApJ, 807, 2

Froment, C., Auchère, F., Aulanier, G., et al. 2017, ApJ, 835, 272

Harra, L. K., Sakao, T., Mandrini, C. H., et al. 2008, ApJL, 676, 2

Higginson, A. K. et al. 2016, ApJ, 837, 2

Kawaguchi, I. 1970, PASJ, 22, 405

Kayshap, P., Banerjee, D., & Srivastava, A. K. 2015. SoPh 290, 10

Klimchuk, J. A., Karpen, J. T., & Antiochos, S. K. 2010, ApJ, 714, 2

Lau, Y., & Finn, J. M. 1990, ApJ, 350, 1

Li, L., Zhang, J., Peter, H., Chitta, L. P., Su, J., et al. 2018, ApJL 864, 1

Liu, W., De Pontieu, B., Vial, J. C., et al. 2015, ApJ, 803, 2

Luhmann, J. G., Li, Y., Zhao, X., et al. 2013, SoPh, 213, 367

Mackay, D. H., Devore, C. R., Antiochos, S. K., et al. 2018, ApJ, 869, 62

Mok, Y., Drake, J. F., Schnak, D. D., et al. 1990, ApJ, 359

Müller, D. A. N., De Groof, A., De Pontieu, B., et al. 2005, ESA SP, 667

Müller, D., Nicula, B., Felix, S., et al. 2017, A&A, 606

O'Shea, E., Banerjee, D., & Doyle, J. G. 2007, A&A, 28, 25

Pariat, E., Antiochos, S. K., & Devore, C. R. 2009, ApJ, 691, 61

Parnell, S. J., & Edwards, C. E. 2015, SoPh, 290, 2055

Raouafi, N.-E., Petrie, G. J. D., Norton, A. A., et al. 2008, ApJL, 682, L137

Raouafi, N.-E., Patsourakos, S., Pariat, E., et al. 2016, SSRv, 201, 1

Schrijver, C. J. 2001, SoPh, 198, 325

Scullion, E., Wedemeyer, S., & Antolin, P. 2014, ApJ, 797, 36

Scullion, E., Van Der Voort, L. R., Antolin, P., et al. 2016, ApJ, 833, 2

Sheeley, Jr., Martin, S.F., Panasenco, O., et al. 1975, SoPh, 40, 103

Shibata, K., Ishido, Y., Acton, L. W., et al. 1992, PASJ, 44, L173

Shibata, K., Nitta, N., Strong, K.T., et al. 1994, ApJL, 431, L51

Titov, V. S., Linker, J. A., Lionello, R., et al. 2011, ApJ, 731, 111

Tousey, R., Bartoe, J.-D. F., Bohlin, J. D., et al. 1973, SoPh, 33, 265

Ugarte-Urra, I., Warren, H. P, & Winebarger, A. R. 2007, ApJ, 662, 1293

Vourlidas, A., & Bastian, T. S. 1996, SoPh, 163, 99

Wang, Y.-M. 1994, ApJ, 435, L153

Wyper, P. F., Antiochos, S. K., & DeVore, C. R. 2017, Natur, 544, 452

Yang, K., Guo, Y., & Ding, M. D. 2015, ApJ, 806, 171

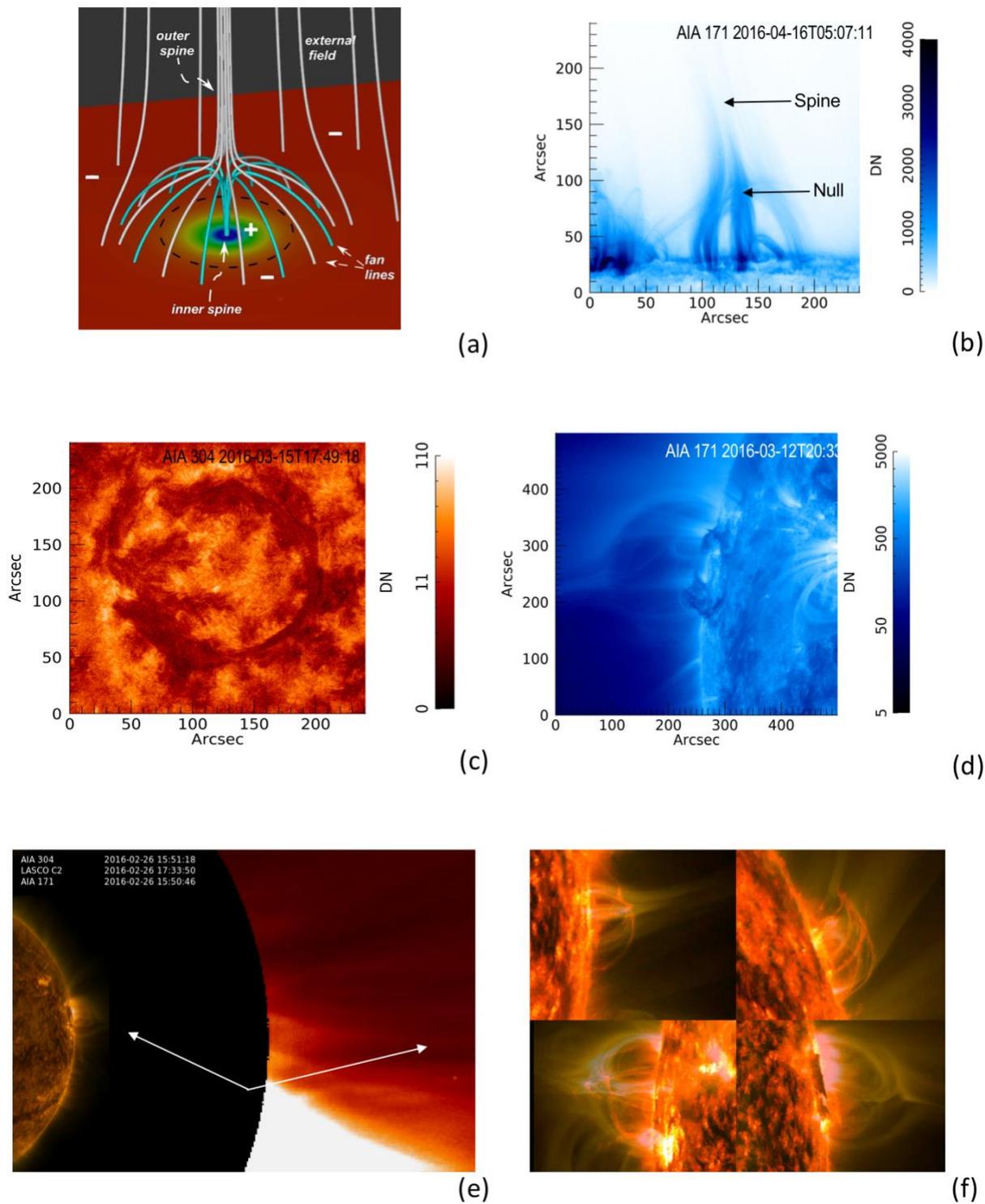

Figure 1: (a) Diagram of the magnetic topology here named as RNPTs (Figure 1 in Pariat et al. 2009). (b) False-colored AIA 171 Å image showing the coronal signature of an RNPT. (c) and (d) examples of the shape and location in 304 and 171 Å wavelengths, respectively, of a near-circular prominence beneath the dome of Event 3. (e) We use white arrows to indicate the apparent location of the spine from Event 3 extending out beyond 5 R$_s$ in LASCO imagery (Helioviewer as described in Müller et al. 2017). (f) We show still images from portions of Movie 1, online (32 s); the movie shows long periods of coronal rain in each of the three events covered in this Letter, using AIA 171 and 304 Å imagery. Event 3 (see section 2.2.3) is covered on three separate dates, to show how the rain formation rate drops as the RNPT ages.

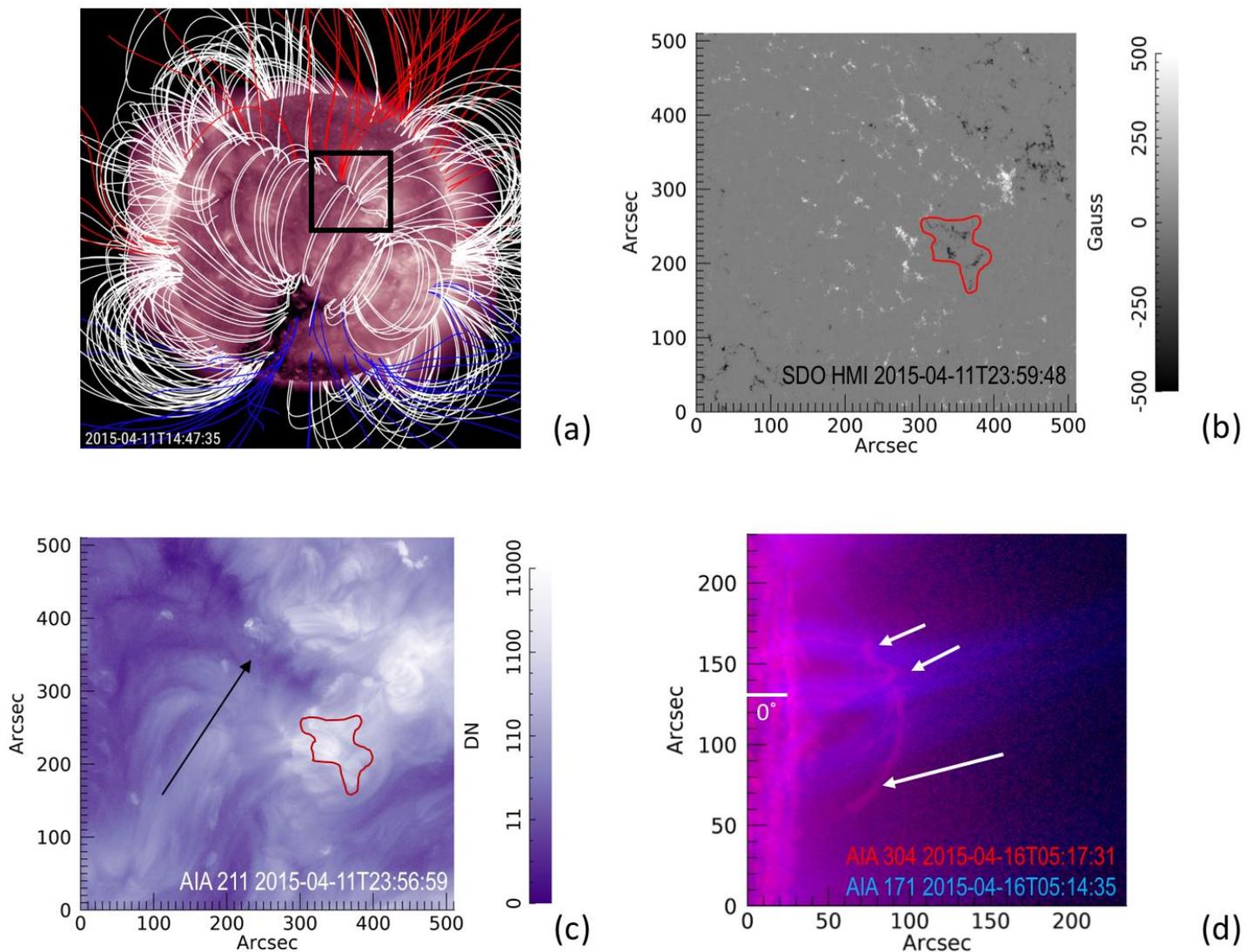

Figure 2: (a) Context image of the Sun on April 11th, including PFSS overlay (open field lines are in red and blue). We show the field of view for the next two images as a black rectangle. (b) SDO HMI data from April 11th, when the magnetic structure underlying the RNPT was more clearly visible on-disk. We delineate the approximate location of the polarity inversion line with the red closed curve. (c) Same region in 211 Å, also on April 11th for coronal context and to help show the coronal hole: the PIL is also overlaid on this image. We use the black arrow to draw attention to the coronal hole extension that borders the RNPT. (d) RNPT on the limb at the solar equator; we show AIA 171 in blue, and AIA 304 Å in red.

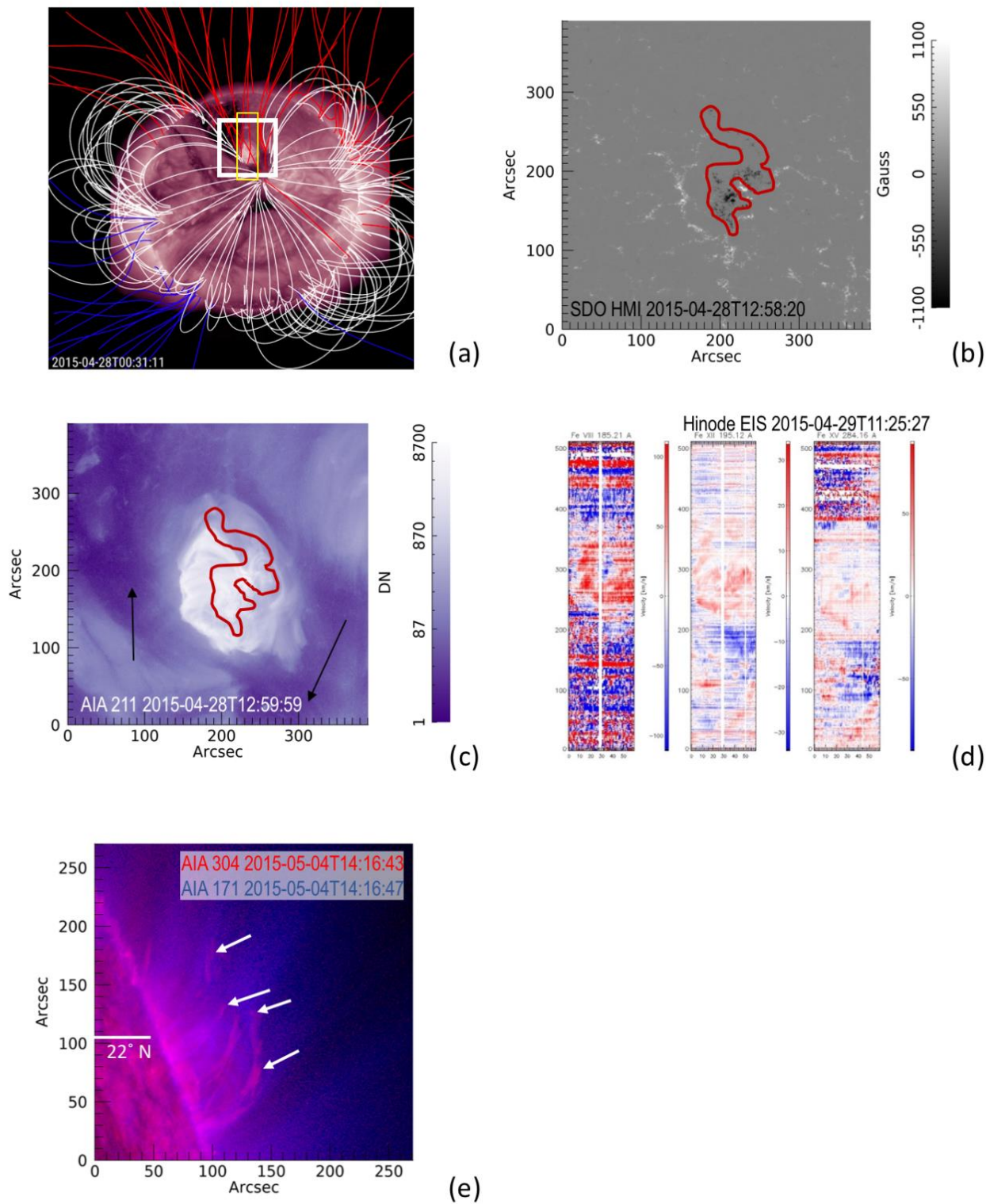

Figure 3: (a) Context image of the Sun on April 28th in 211 Å, with the field of view of the subsequent two images overlaid in white, and the field of view of Hinode/EIS observations in yellow. (b) Magnetic field on April 28th, observed by SDO several days before the RNPT appears on the limb at 24° N. The red line delineates the rough location of the polarity inversion line. (c) AIA 171 Å image of the RNPT embedded in an extension of the polar coronal hole (indicated by the black arrows). (d) Series of EIS Doppler maps of the RNPT and surrounding open field regions in a range of lines. These lines have peak formation temperatures around 0.5, 1.6, and 2.2 MK respectively. e) AIA 171 and 304 Å imagery of rain in Event 2.

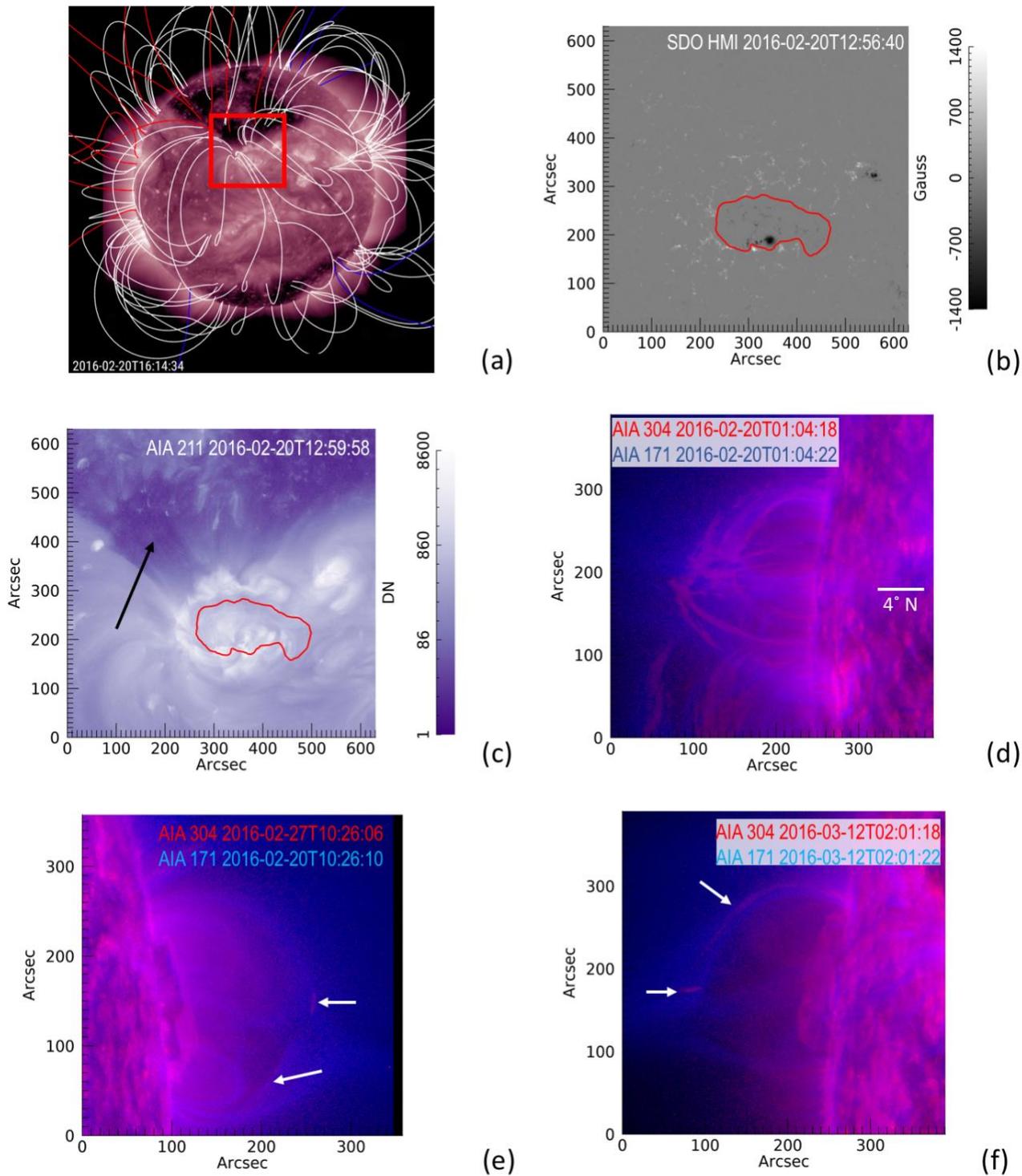

Figure 4: (a) Context image in 211 Å, with the field of view for the next two panels highlighted in red. (b) HMI imagery on February 20 for AR 12333, analogous to that in the previous figures. (c) RNPT and the surrounding coronal signatures in 211 Å. As in the preceding figures, we have overlaid the PIL on the image, and the black arrow indicates the coronal hole. (d) We exhibit the closed dome structure in the corona and highlight some of the rain that fell on February 15th. (e) AIA 171 and 304 Å images of the RNPT on the west limb on 2016 February 27; the white arrows show the coronal rain. (f) AIA 171 and 304 Å imagery of the RNPT on the east limb on 2016 March 12.

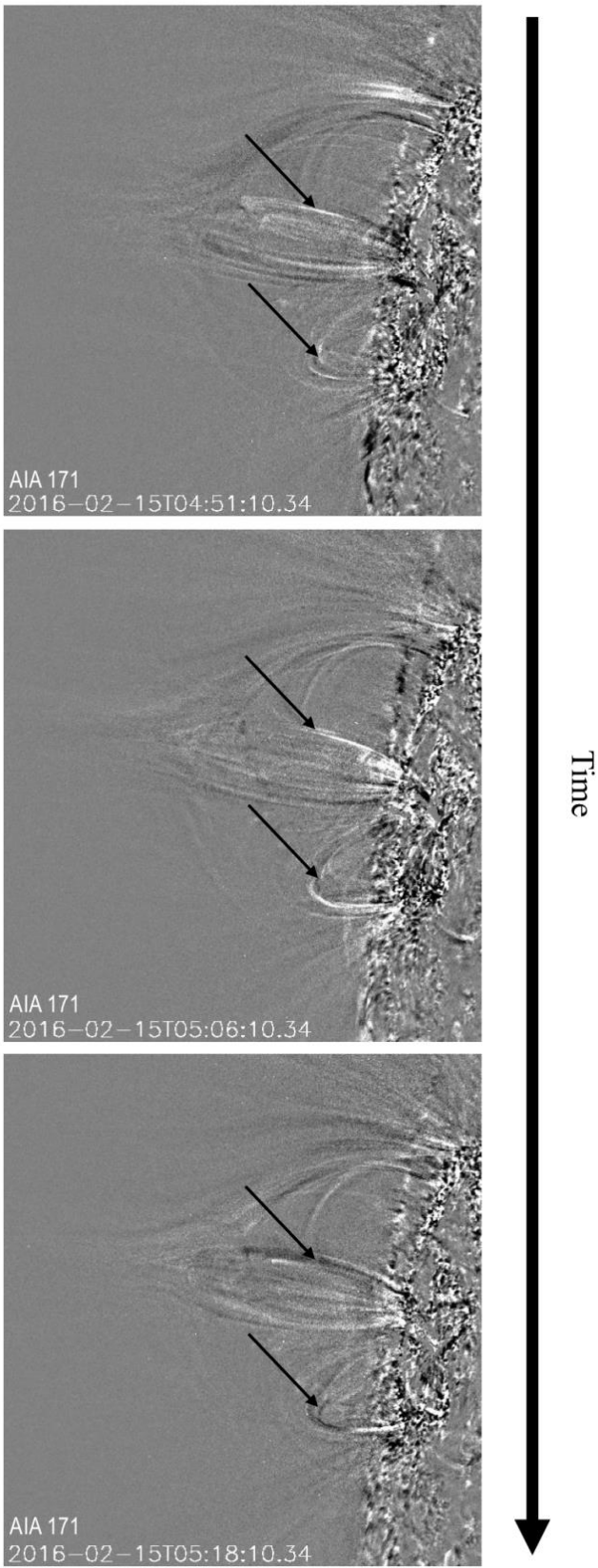

Figure 5: Processed time-difference frames of AIA 171 Å, with two black arrows pointing out white, distinct blob intensifications. As time progresses, the blobs fall and are replaced with dark tracks in the later difference images. The arrows point at the same locations in all three images for ease of continuity. (Credit Nathalia Alzate, using family of techniques outlined in Plowman 2016)